\documentclass[%
 preprint,
superscriptaddress,
 amsmath,amssymb,
 aps,
]{revtex4-2}

\usepackage{graphicx}
\usepackage{dcolumn}
\usepackage{bm}
\usepackage{xcolor} 

\usepackage[version=3]{mhchem} 

\usepackage{amsmath,amssymb}
\usepackage{amsfonts}
\usepackage{stackengine}
\stackMath
\usepackage{natbib}
\usepackage{array}
\usepackage{amssymb}
\usepackage{multirow}
\usepackage{mathtools}
\newcolumntype{z}[1]{D{.}{.}{#1}}
\usepackage{hyperref}
\usepackage{float}
\usepackage{mciteplus}

\usepackage{tabularx}
\usepackage{rotating}
\usepackage{placeins}
\usepackage{romannum}
\hypersetup{
    colorlinks=true,
    linkcolor=black,
    citecolor=black,
    filecolor=black,
    urlcolor=black,
}

\usepackage{physics}
\usepackage{soul,xcolor}

\begin{document}

\setstcolor{red}
\preprint{APS/123-QED}

\title{From Branched Flow to Superwires in Periodic Potentials}

\author{Alvar Daza}
 \affiliation {Department of Physics, Harvard University, Cambridge, Massachusetts 02138, USA}
 \affiliation{Nonlinear Dynamics, Chaos and Complex Systems Group, Departamento de F\'{i}sica, Universidad Rey Juan Carlos, M\'{o}stoles, Madrid, Tulip\'{a}n s/n, 28933, Spain}
\email{alvar.daza@urjc.es}

\author{Eric J. Heller}
 \affiliation {Department of Physics, Harvard University, Cambridge, Massachusetts 02138, USA}
 \affiliation {Department of Chemistry and Chemical Biology, Harvard University, Cambridge, Massachusetts 02138, USA}

\author{Anton M. Graf}
\affiliation {Harvard John A. Paulson School of Engineering and Applied Sciences, Harvard, Cambridge, Massachusetts 02138, USA}

\author{Esa R\"as\"anen}
\affiliation{Computational Physics Laboratory, Tampere University, Tampere 33720, Finland}
\affiliation {Department of Physics, Harvard University, Cambridge, Massachusetts 02138, USA}
\email{esa.rasanen@tuni.fi}

\date{\today}

\begin{abstract}
We report unexpected classical and quantum dynamics of a wave propagating in a periodic potential in high Brilloiun zones. Branched flow appears at wavelengths shorter than the typical length scale of the ordered periodic structure and for energies above the potential barrier. The strongest branches remain stable indefinitely and may create linear {\it dynamical} channels, wherein waves are not  confined directly by potential walls as electrons in ordinary wires, but rather indirectly and more subtly by dynamical stability. We term these superwires, since they are associated with a superlattice.
\end{abstract}

\maketitle

\section{Introduction}
Branched flow is a common phenomenon of wave dynamics: when a wave impinges a weakly refractive medium, it can create an intensity pattern akin to the shadow of a tree~\cite{heller2019branched}. Unlike normal diffusion, some of the (temporarily and accidentally) stable branches can carry a high density of flux across long distances. Branched flow is important
on hugely disparate scales, from electron waves in two-dimensional electron gas~\cite{topinka2001coherent}, to acoustic waves spanning thousands of kilometers in the oceans~\cite{degueldre2016random}, or the beautiful patterns of light going through soap bubbles~\cite{patsyk2020observation}. All these phenomena in both classical and quantum systems 
arise from wave propagation in {\em random} potentials.

 As a general rule, branched flow for waves appears when the wavelength $\lambda_F$ is shorter than the typical length scale $a$ of the potential given small angle deflections per "feature" in the potential.
 In most materials the lattice constants are of the order of Angstroms, whereas the electron wavelengths are in the nanometer scale, i.e., $\lambda_F > a$, so we may expect that branched flow cannot exist
 in crystals. However, in the last years there have been significant research activities on superlattices, where the combined periodic structures may create a larger-scale periodic structure. A perfect example is twisted bilayer graphene that exhibits a large-scale moiré pattern~\cite{bistritzer2011} and exotic properties such as superconductivity~\cite{cao2018unconventional,cao2018b,yankowitz2019,lu2019,stepanov2020}. 
 As the condition $\lambda_F < a$ in these superlattices is generally
 satisfied, we may expect that branched flow -- if it exists despite the periodicity -- provides important understanding on the physical properties of novel "designer materials" including layered structures~\cite{kim2017,chen2019,tartakovskii2019}, artificial lattices~\cite{gomes2012,polini2013,slot2017,Khajetoorians2019}, and photonic systems~\cite{ReserbatPlantey2021}.

In this work we extend the concept of branched flow to periodic potentials. Thus, we demonstrate the ubiquity of branched flow from classical and quantum scales and from random disorder to periodic systems. But perhaps even more important than these irregular patterns, are the indefinitely stable branches that can arise  in periodic potentials. Within these controlled branches, propagating waves are dynamically confined, creating superwires. Unlike wires based on energetic barriers, these superwires arise because of the dynamics. In this regime, waves could surmount the potential barrier, but their dynamics keep them in a narrow spatial region.

 The paper is organized as follows. In Sec.~\ref{sec:methods}, we outline the computational methods employed in the study of branched flow in both classical and quantum regime, which share many features in the semiclassical limit. In Sec.~\ref{sec:shock} we demonstrate the appearance and properties of branched flow in periodic systems compared to the conventional branched flow and the Bloch wave representation. Further, in Sec.~\ref{sec:Branched} we examine the classical picture that provides insight about the origin of branched flow and its relation to chaos. Channeling effects in terms of long-lived stable branches are studied in detail in Sec.~\ref{sec:Channeling}. Finally, the possible implications of our findings and the future directions are discussed in Sec.~\ref{sec:Discussion}

\section{Methods}
\label{sec:methods}

Branched  flow is typically examined by classical   trajectories,  and by  time dependent wave packet calculations under the influence of random potentials~\cite{heller2018semiclassical}. For periodic superlattices,  we also use both classical and wave packet analysis, finding both branched and superwire flow. The 2D  results are supplemented by the simpler 1D ``kick and drift'' map,  aiding understanding of   branching and  superwire dynamical channeling.

 The evolution of the wave packet is computed using the split-operator technique~\cite{feit1982solution}. This iterative method comprises several steps: (i) the initial state evolves under the action of the potential in the coordinate representation $\Psi \rightarrow e^{-i V(q)\tau / \hbar} \Psi$, (ii) the resulting state is Fourier transformed into momentum representation $\Psi \rightarrow \hat{\Psi}$, (iii) the state evolves in the momentum representation $\hat{\Psi} \rightarrow e^{-i p^2\tau / 2m\hbar} \hat{\Psi}$, (iv) an inverse Fourier transform gives back the resulting state to the coordinate representation $\hat{\Psi}(t+\tau) \rightarrow \Psi(t+\tau)$. This procedure provides a fast and reliable method to study the wave dynamics, as long as the time step $\tau$ is small. In fact, most of the pictures depicted here were computed within a few minutes in a regular workstation.

If the quantum wavelength is short enough, the semiclassical approach will be valid,
making a classical analysis very informative even though the goal is to understand quantum systems. We study the density of a large ensemble (typically thousands) of classical trajectories, using initial distributions analogous to the quantum ones. Given the Hamiltonian nature of the problem, integration is carried out with a symplectic scheme \cite{verlet1967computer,yoshida1993recent} preserving the phase space volume and  the energy in all cases. We employ a computational cluster to perform the classical simulations in reasonable times.

\section{Branched flow in periodic systems}
\label{sec:shock}

\begin{figure}[!h]
    \centering
    \includegraphics[width=\linewidth]{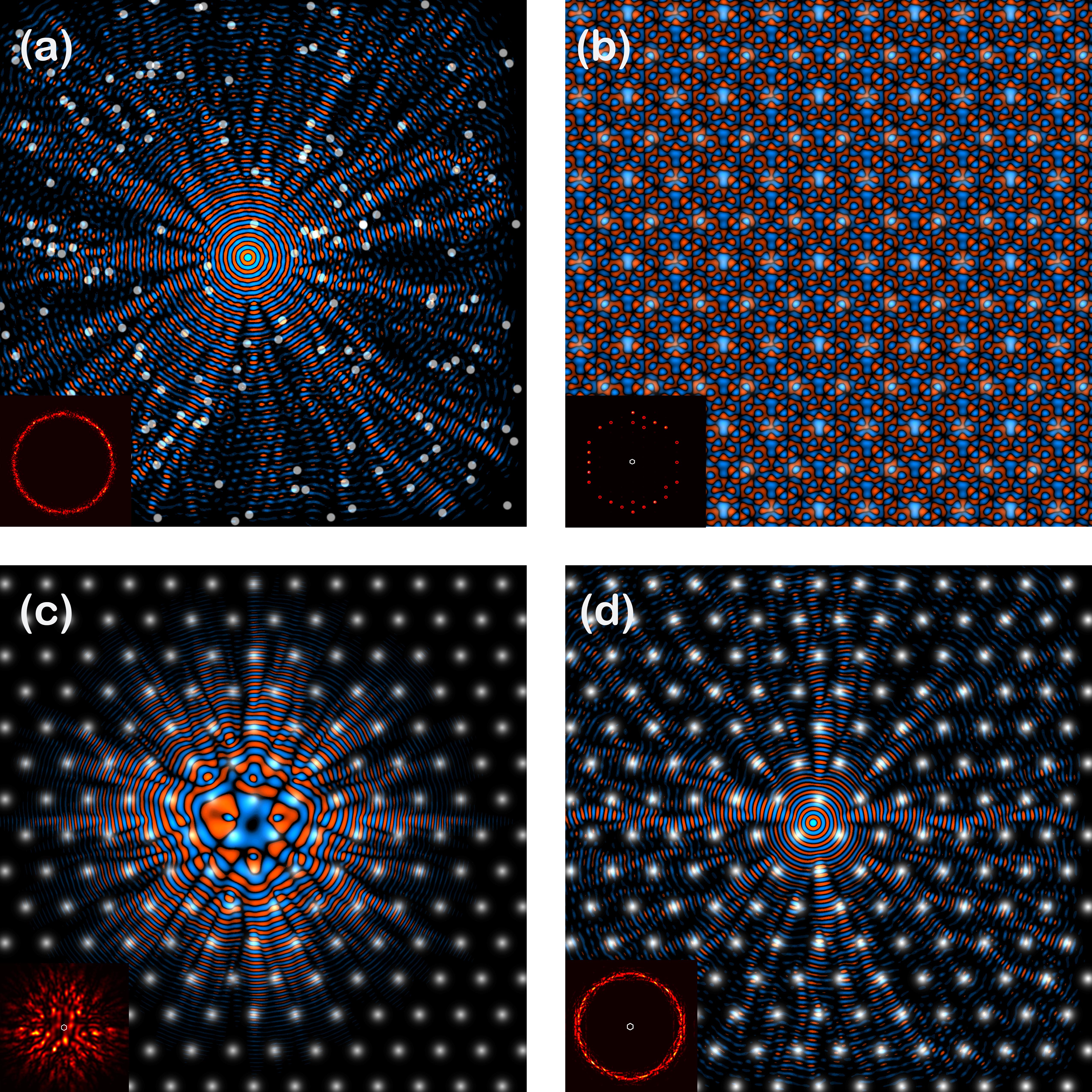}
    \caption{(a) Branched flow in a potential made out of randomly spaced wells. (b) Bloch wave ($\vec{k}=\vec{0}$) for a triangular potential. (c) Snapshot of a Gaussian wave packet evolving in a triangular superlattice. The real part of the wave is depicted, with red for the positive and blue for the negative parts. The fast components lead the evolution of the wave on the periphery, while the slower components lag near the center. (d) Eigenfunction of the wave packet of (c), made using an energy Fourier transform. The insets show the momentum representation of the corresponding panels.}
    \label{fig:mag}
\end{figure}

Figure~\ref{fig:mag}(a) shows conventional branched flow in a two-dimensional potential characterized by randomly positioned wells (gray dots). The initial state corresponds to a narrow Gaussian wave packet localized at the center. The wells are modeled by soft Fermi-type potentials (see Ref.~\cite{lorentzgas} and below) with an amplitude that corresponds to half of the energy of the wave packet. The characteristics of the branching produced by this random potential are similar to the previous findings~\cite{heller2019branched}. 


In Figs.~\ref{fig:mag}(b-d) the potential is similar to (a), but the wells are arranged to a periodic triangular lattice. Intuitively, we may expect the system to be characterized by Bloch waves -- see Fig.~\ref{fig:mag}(b) as an example. However, the propagation of the wave packet under the periodic potential leads to branched flow that is astonishingly similar to conventional branched flow; see Figs.~\ref{fig:mag}(c-d) that are discussed in detail below. The insets show the momentum Fourier transform of the corresponding figures. This representation does not correspond exactly to the reciprocal space (we would need to make Bloch wave Fourier transformation), but it helps to understand the different regimes.

First, Fig.~\ref{fig:mag}(c) shows a snapshot of the full wave packet during the evolution. The components with short wavelengths propagate faster, whereas the components with longer wavelengths lag behind and remain closer to the origin. As the wave evolves, its Fourier transform $ \Psi_E=\int_{-\infty}^{\infty} e^{-i E t /\hbar} dt$ for some particular energy $E$ can be accumulated. For sufficiently large lattices, waves exiting the observation window never return. This means that the integral over infinite time can be reduced to the observation time that each component of the wave packet takes to leave the picture. By using an absorbing potential around the observation window the filtered state $\Psi_E$ is an eigenfunction by construction, but its morphology can be very complicated without spatial periodicity as in the Bloch wave.

Figure~\ref{fig:mag}(d) shows an eigenfunction of the wave packet in (c). Chaotic branching similar to conventional branched flow in (a) is clearly visible. Some branches are localized on top of the bumps whereas others are avoiding them. This can be a hint for the presence of quantum scars  \cite{kaplan1999scars,luukko2016strong, keski2019quantum}. Classically, these regions correspond to unstable periodic orbits of chaotic systems. Such unstable trajectories belong to a set of measure zero in the classical picture, but surprisingly the probability of the quantum wave is enhanced in these regions.

As another feature in Fig.~\ref{fig:mag}, several wavelengths can fit in between consecutive bumps. This indicates that these branched eigenfunctions live far beyond the first Brillouin zone. Of course, everything can be folded into the first Brillouin zone, but we might lose some intuition by such operation. Nevertheless, the property $\lambda_F < a$ discussed above is clearly fulfilled, and it has direct relevance for, e.g., twisted bilayer graphene and other moir{\'e} superlattices~\cite{bistritzer2011,cao2018unconventional,cao2018b,yankowitz2019,lu2019,stepanov2020}.

\begin{figure}[!h]
\includegraphics[width=\linewidth]{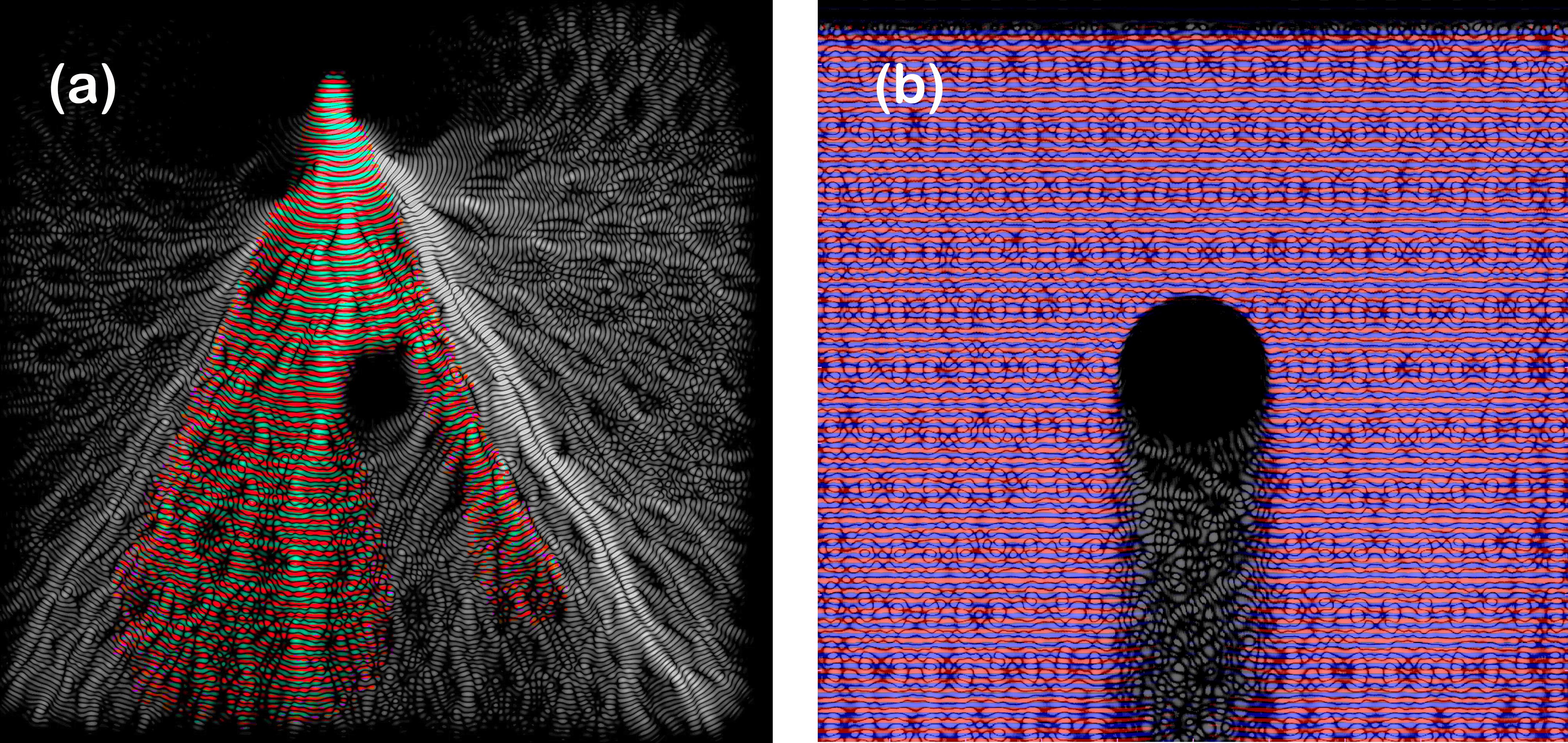} 
\caption{(a) Wave packet propagation with downwards initial momentum in a triangular potential (grayscale) exposed to an absorbing disk at the center of the figure. The colors correspond to the same propagation but in free space without the triangular lattice. (b) Same as in (a) but now the initial state is a Bloch wave. The absorbing disk casts a shadow for the free wave, but this space is filled by branches when the triangular potential is present. 
}
\label{fig:Comparison}
 \end{figure}

Next we focus on the role of the periodic potential in branching and on the complementarity between branched eigenfunctions and Bloch waves. Figure~\ref{fig:Comparison} shows a comparison between the evolution of a wave in a periodic triangular potential (grayscale) and in free space without the potential (colorscale). The black circle in the middle corresponds to an absorbing potential. In Fig.~\ref{fig:Comparison}(a) the initial state is a wave packet with downward momentum, whereas in (b) the initial state is a Bloch wave which is also descending. In both cases, the free plane wave casts a hard shadow (no colors) as it passes the absorbing hole. Instead, the periodic potential causes clear branches (grayscale) behind the disk. In the case of the Bloch wave in Fig.~\ref{fig:Comparison}(b), the periodic potential thus filters out the underlying branched fabric.

Figure~\ref{fig:Comparison}(a) shows further surprising effects. First, the periodic potential eventually leads to backward propagation, which is prominent in the upper-right corner. In the same region, we can also clearly see the periodic structure resulting from the triangular potential. Such regular patterns may emerge in a certain wavelength range, even though the majority of the evolution occurs in a randomized fashion. Finally, on the right side of the colored region we find a relatively straight branch in light gray that has a longitudinal node in between. This is reminiscent of a dynamical channeling effect analyzed in detail in Sec.~\ref{sec:Channeling}.

\section{Dynamics of branched flow}
\label{sec:Branched}

\subsection{Integrable and non-integrable potentials}

Even though the results above show that branched flow can be found in both randomized and periodic potentials, we remind that 
the effect takes place only for sufficiently high energies compared to the underlying potential~\cite{heller2018semiclassical}. Otherwise, the flow is trapped in the troughs of the potential and the dynamics is characterized by different types of classical {\em diffusion}, (sub-, normal, super-, anomalous) and Lévy flight behavior depending on the system parameters~\cite{Cristadoro2014,Zarfaty2018,lorentzgas}.
Furthermore, branched flow also requires the wavelength to be sufficiently small compared to the scale of the potential. As discussed above, this makes branched flow in superlattices relevant, given the relation between typical electron wavelengths and superlattice spacing. If the wavelength is comparable or even larger than the lattice spacing, the wave ignores the potential just as light becomes transparent through window glass. However, besides these requirements, there is another important ingredient that has not been explicitly studied before, i.e., the integrability of the potential.

\begin{figure}
\includegraphics[width=0.9\linewidth]{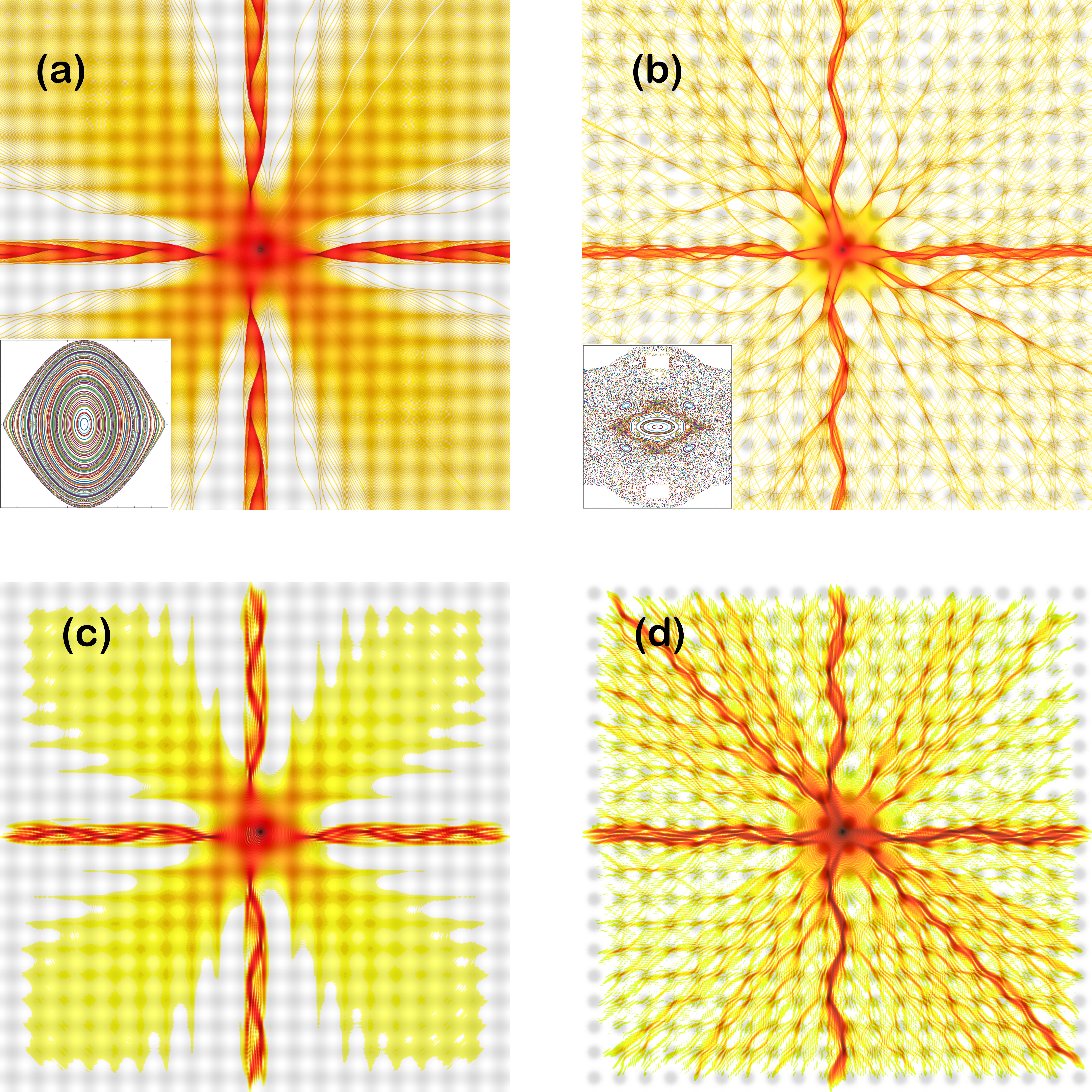}
\caption{(a) Density of classical trajectories in a square lattice defined by an integrable potential. (b) Same as (a) but in a non-integrable potential. (c) Density of a quantum wave function in an integrable potential. (d) Same as (c) but in a non-integrable potential.
}
\label{fig:ChaosIntegrable}

 \end{figure}

In Fig.~\ref{fig:ChaosIntegrable} we compare the dynamics of the wave packet in both classical (a-b) and quantum (c-d) simulations using both integrable (a,c) and non-integrable (b,d) potentials. The integrable potential is defined as $V=-A(\cos x + \cos y)$, which corresponds to a square lattice that is revealed by the darker regions of the picture. For this potential, the motion can be separated into $x$ and $y$ in terms of Jacobi elliptic functions. The density of classical trajectories in Fig.~\ref{fig:ChaosIntegrable}(a) shows focusing of the beam along the four main channels, but the pattern is repetitive and predictable. Once a trajectory enters one of these channels, it remains confined within its narrow boundaries as long as the potential remains periodic. This behavior is examined in more detail in Sec.~\ref{sec:Channeling}.

In comparison, Fig.~\ref{fig:ChaosIntegrable}(b) shows the density of classical trajectories in a {\em non-integrable} Fermi-type potential defined as $V(\vec{r})=\sum_{j=1}^N A/\left[1+\exp(|\vec{r}-\vec{r_{0j}}|/\sigma)\right]$, which is also used above within Figs. 1 and 2. Now, $\vec{r_{0j}}$ provides the location of each of the $N$ bumps of a square lattice (gray dots). In this case, a more intricate pattern emerges, including branches showing up at non-trivial locations and carrying a high density of flow along variable lengths. The drastic difference from an integrable case in Fig.~\ref{fig:ChaosIntegrable}(a) shows that the non-integrability -- corresponding to a chaotic system -- is the key ingredient behind branched flow. This is demonstrated also by the phase-space pictures given as insets: foci arise as a consequence of cusp catastrophes, which can occur in integrable and chaotic dynamics, but the distribution and stability of these foci is much richer when the phase space is scrambled, corresponding to a branched flow.

In Figs.~\ref{fig:ChaosIntegrable}(c) and (d) we show the corresponding densities of {\em quantum} wave functions evolved under the same integrable and non-integrable potentials, respectively. The agreement between the classical and quantum simulations is evident, as well as the differences between the integrable and non-integrable potentials. Hence, these results demonstrate that the main ingredients required for the phenomenon of branched flow are similar both classically and quantum mechanically. In particular, the non-integrability of the potential is a necessary condition -- and also the default condition from an experimental perspective. Furthermore, the characteristics of the main and secondary branches in classical and quantum cases are very much alike, as clearly seen in Fig.~\ref{fig:ChaosIntegrable}.

\subsection{Kick and drift map}

To analyze the  branched flow in a periodic system further, let us consider the classical kick and drift map~\cite{heller2019branched}. It is an area preserving time-discrete map based on Hamilton's equation of motion defined by
\begin{equation}
 \begin{split}
    \vec{p}_{n+1} &= \vec{p}_n - \overrightarrow{\nabla V}\vert_{\vec{x}=\vec{x}_n},\\
    \vec{x}_{n+1} &= \vec{x}_n + \vec{p}_{n+1},
     \end{split}   
\end{equation}
where $\vec{x}$ and $\vec{p}$ correspond to the trajectory's position and momentum, $n$ is a natural number playing the role of discrete time, and $V$ is a potential depending on the position. The kick and drift map receives its name because of its two stages: first the momentum changes according to the potential, and then the trajectory drifts until the next kick. This simple picture provides useful insights about the phase space transformations that give rise to branched flow, including the creation of foci through cusps and the stability of the long-lived branches, among other interesting effects. 

 \begin{figure*} 
\includegraphics[width=\textwidth]{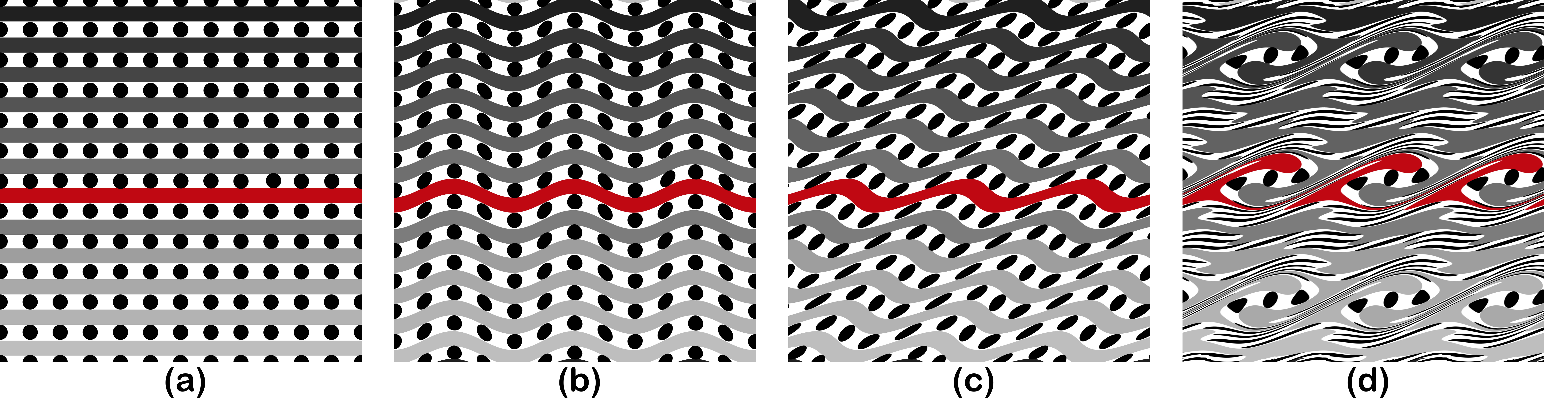}
\caption{(a) Stripes and circles represent different initial manifolds in phase space $(x,p)$. A periodic kick in momentum produced by the potential, and a free drift (shear in phase space), transform the initial sketch of (a) into (b) and (c) respectively. (d) After a few kicks and drifts, the initial manifolds have been stretched and folded giving rise to cusp catastrophes. Also, some circles remain almost unaltered, which would correspond to stable regions in phase space.
}
\label{fig:sequence}
 \end{figure*}

Previously,~\cite{heller2019branched}, the "kick" used randomly chosen parameters, but here we repeatedly use the same spatially periodic potential, writing
\begin{equation}
 \begin{split}
    {p}_{n+1} &= {p}_n + K{\sin x_n},\\
    {x}_{n+1} &= {x}_n + {p}_{n+1},
\label{eq:Standard}
     \end{split}   
\end{equation}
where $K$ accounts for the height of the potential. Equations~\ref{eq:Standard}  define the celebrated standard map, studied by Boris Chirikov in the context of Hamiltonian chaos and Kolmogorov-Arnold-Moser (KAM) theory \cite{chirikov1969research, chirikov1979universal}. For low values of the perturbation strength $K$, periodic motion dominates, whereas for higher values, the phase space  increasingly fills with chaotic trajectories. Stable islands in the standard map correspond to channels or superwires in a lattice, and branched flow corresponds to the predominantly unstable standard map zones. We can see the effect of the standard map on different manifolds on~Fig.~\ref{fig:sequence}. In this phase space sketch, horizontal stripes of panel (a) would correspond to plane waves (different positions, same speed). The periodic kick and drift of Eq.~\ref{eq:Standard} reshape these manifolds, as shown in~Fig.~\ref{fig:sequence}(b)-(c). After a few steps, we can see red blobs indicative of the cusp formation. Also, some of the circles remain almost unaltered, which would correspond to stable regions in phase space, and potentially superwires.

However, there is a crucial difference of perspective compared to most of the standard map-chaos literature: we focus on the  early and medium time development, or temporal evolution, of the phase space structure, as in the study of branched flow.

\section{Stability and superwires}
\label{sec:Channeling}

According to our results above, branched flow is produced when a classical or quantum wave with sufficient energy impinges on potential landscape (periodic or not) -- as long as the potential is not integrable. By examining Fig.~\ref{fig:ChaosIntegrable}, we can see that the four arms of the cross both in the integrable and the non-integrable case remain stable for long times. Here we will refer to these regions as channels -- or superwires (cf. superlattices) -- of the flow that remain bounded for long times.

\begin{figure*} 
\includegraphics[width=\textwidth]{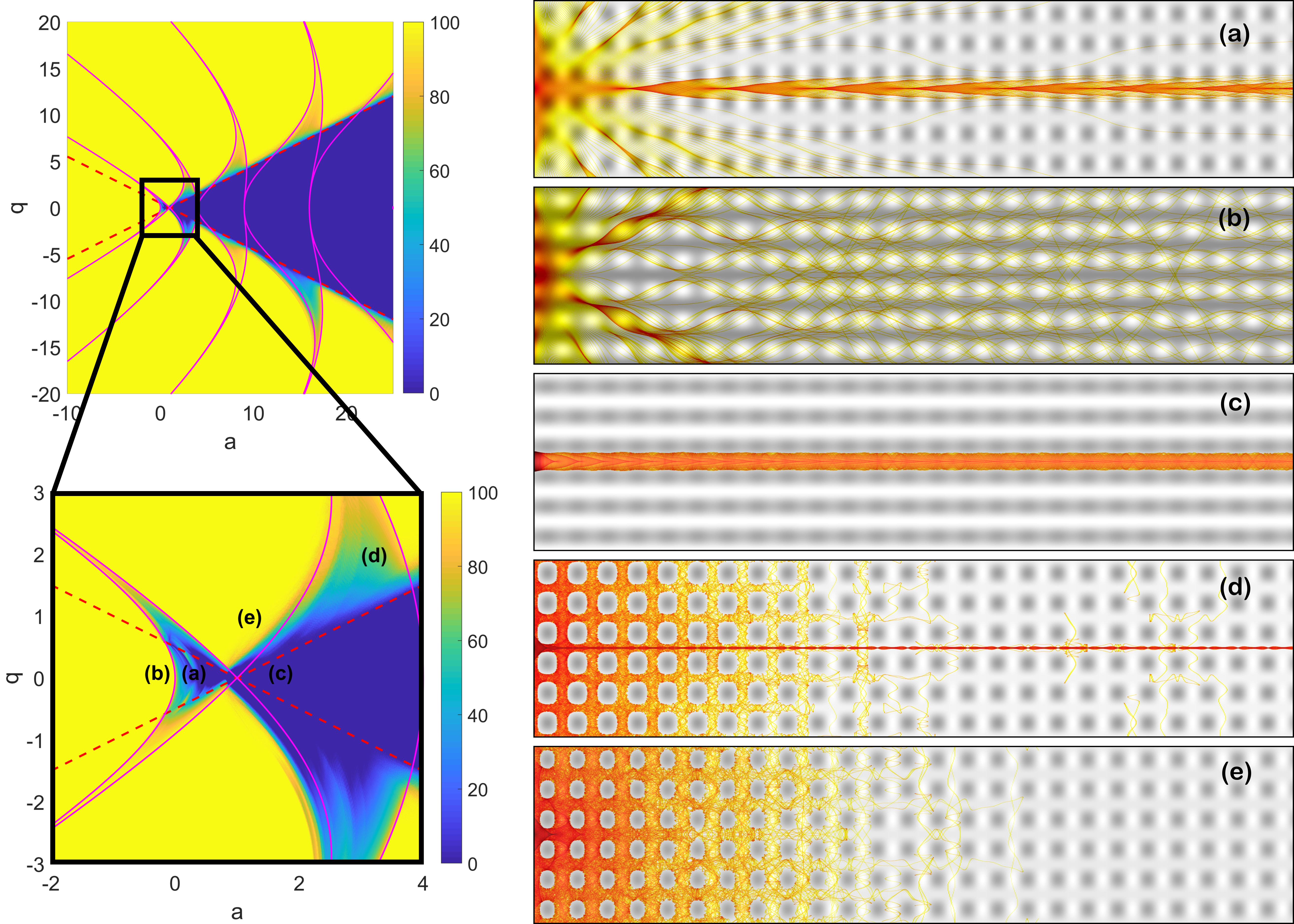}
\caption{On the left, the stability diagram for the potential of Eq.~\ref{eq:VMathieu}. The color code shows the percentage of trajectories that remain within the initial channel from an initial wedge spanning $60^\circ$. The magenta lines are for the stability of Eq.~\ref{eq:Mathieu}, while the red dashed lines are for $q=(1-a)/2$ and $q=(a-1)/2$, to help identify the different regions in terms of energy. (a) Classical simulation of a superwire, where trajectories remain confined due to the dynamics. For parameters in (b), the horizontal channel is not populated. (c) Energetically confining channel: trajectories cannot surmount the barriers. (d) Trajectories can escape in between the bumps, but cannot override them. The dynamical confinement is not as strong as the one in (a). (e) Trajectories explore the potential chaotically. 
}
\label{fig:Mathieu}
 \end{figure*}

The stability of the  channels can be understood in terms of motion normal to the superwire paths, which can be approximated by  Mathieu functions~\cite{mathieu1868memoire, mclachlan1951theory}. Consider a classical trajectory starting in the center of a square lattice of repulsive soft pillars, heading to the right between the rows of bumps (see the right panels of Fig.~\ref{fig:Mathieu} for an example). As the trajectory progresses,  its motion can be linearized around the exact, straight line path down the bumpy rows. By symmetry, the path has no transverse force on it and it remains straight. A stability analysis is needed to determine the fate of nearby trajectories. If they are stable, there are superwire paths oscillating down the row. Expanding the potential to second order normal to the path, the effective potential is a  harmonic oscillator with a force constant that is varying (nearly) periodically. If that variation is approximately sinusoidal, the stability can be assessed with the Mathieu equation,
\begin{align}
      \dfrac{d^2x}{dt^2}+(a-2q\cos{2t})x=0. 
  \label{eq:Mathieu}
\end{align}
The stability of the solutions of this equation has been thoroughly studied~\cite{mclachlan1951theory}. If the time variation has strong harmonics, the analysis and the results are similar, so we are content with the Mathieu analysis for now. 

The key to the stability is the period $\pi$ of the oscillation of $2q\cos{2t}$ relative to the time averaged frequency $\sqrt a$. A channel trajectory is equivalently parameterized by a fixed $a$, given by the shape of the  potential normal to the path. Thus, we can write Eq.~(\ref{eq:Mathieu}) as $d^2x/dt^2+(a-2q\cos{2\omega t})x=0$, where $\omega = 2\pi/\tau$ and $\tau = A/v$ is the time to traverse a unit cell of width $A$ at velocity $v$. The velocity of the flow down the superwire relative to the superlattice parameters becomes crucial. At high velocity, the method of averaging suggests that the trajectory will always become stable. The Mathieu stability diagram confirms this with ever narrowing resonances and larger regions of stability with increasing speed (increasing $\omega$).

Even if the velocity $v$ is not adjustable, as is the case for electrons in twisted bilayer graphene away from the flat band region, the frequency ratio can be controlled by the twist angle, thus adjusting $A$ in $\sqrt a/\omega=\sqrt a/(2 \pi v/A)$.  Or, in artificial superlattices, it can be controlled by fabrication geometry. 
The stability and time dependence of a quantum version of the Mathieu problem is exactly the same as the classical, because it is a harmonic oscillator, a linear dynamical system.  Thus, the classical stability analysis is directly related to the quantum evolution, as confirmed for example in Fig.~\ref{fig:ChaosIntegrable}.
We can test these ideas by constructing a 2D potential with analogous properties to the 1D standard map.  The potential is given by 
\begin{align}
    V=-(2q\cos{2x}-a)\sin{y}^2. 
\label{eq:VMathieu}
\end{align}
Here we have sinusoidal wells instead of harmonic, but by Taylor expansion they are well approximated as harmonic at $y=n\pi$, $n\in \mathbb{Z}$. By computing the evolution of a classical manifold we can see how many trajectories remain within the boundaries of their initial channels after a long time. This is shown in Fig.~\ref{fig:Mathieu}. The magenta lines show the stability lines for Eq.~\ref{eq:Mathieu} in agreement with the simulations. The red dashed lines help to understand the relation between the kinetic and potential energy. Panels (a) and (b) on the right lie on the only part of the $(a,q)$ stability region where trajectories can override the bumps. However, the dynamics of trajectories in (a) keeps them within the channel, thus creating a classical superwire. The periodicity of the focusing of the trajectories is incommensurate with respect to the periodicity of the potential. This is precisely what  makes the channel stable; otherwise the trajectories would be resonant and leave it.

The superwires should not be confused with channels that are trapped energetically, i.e., confined by the bumpy potential. This regime occurs for parameters $(a,q)$ in the region between $q>(T-a)/2$ and $q<(a-T)/2$, where $T$ is the initial kinetic energy (in Fig.~\ref{fig:Mathieu} we have chosen the kinetic energy $T=1$  and this energetically trapping region corresponds to the big blue triangle on the right of the stability plot). As shown in~Fig.~\ref{fig:Mathieu}(c), trajectories in such regime cannot surmount the potential barrier, and consequently they are restricted to a nearly one-dimensional space. Considering a many body problem, this situation could lead to a Luttinger liquid with correlated electrons, but our present study is restricted to the one body problem of a particle in a superlattice. 

Panels (d) and (e) of~Fig.~\ref{fig:Mathieu} show an intermediate regime, where trajectories can escape in between the bumps but cannot ride over the top of them. Some trajectories in these dynamical channels leak out, unlike superwires as the one shown in (a). For such values of the energy, other diffusive mechanisms are at play~\cite{Cristadoro2014,Zarfaty2018,lorentzgas}, hampering controlled transport of the flow.

Different potentials could be built where superwires would be the dominant regime in the parameter space. The potential of Eq.~\ref{eq:VMathieu} is specifically designed to mimic the stability of the Mathieu equation, but for example, variations on the Fermi potential discussed above would also be ruled by the Mathieu equation in the vicinity of the minimum (maximum) between the bumps (wells).

\begin{figure}
\includegraphics[width=\linewidth]{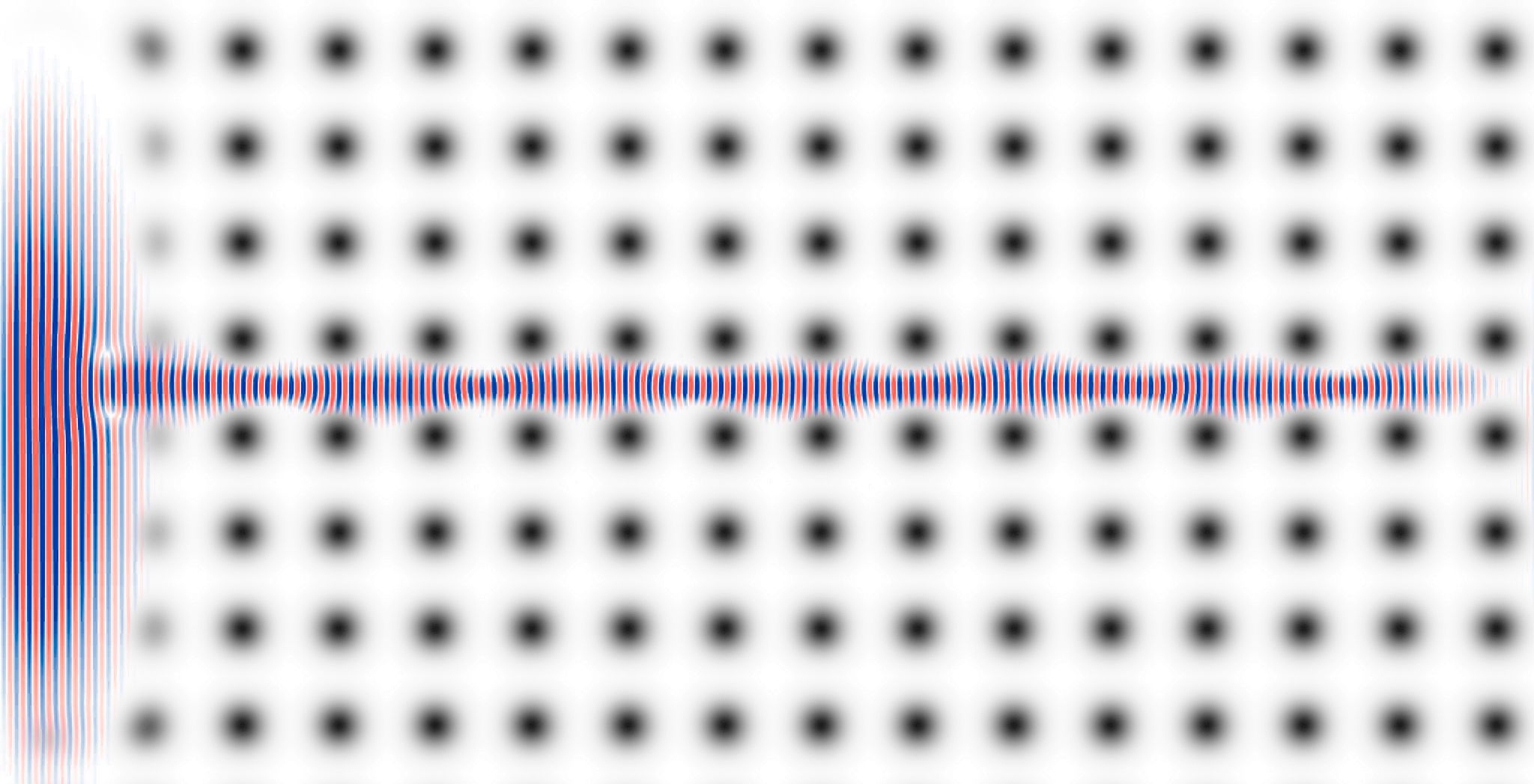}
\caption{Example of a stable superwire. A quantum wave packet is injected from the left into the square lattice. A Fourier transform at a chosen energy reveals a stable superwire along a channel. The potential, mass, and velocity are in a dynamically stable regime. Notice the difference between the periodicity of the propagating wave and the periodicity of the potential. 
}
\label{fig:Channel}

 \end{figure}

Finally we demonstrate the formation of superwires in a quantum mechanical calculation. Figure~\ref{fig:Channel} shows an example of
a wave packet propagation from the left into a square lattice. The wave packet is Fourier-transformed from time to energy at a chosen energy, which reveals a dynamically stable superwire along the channel between the bumps. However, if the lattice were extended further to the right, tunneling to the neighboring parallel channels would eventually occur. This dynamical tunneling~\cite{davis1981quantum} would correspond to the existence of a flat electronic band along the $\vec k$ direction normal to the channel. It is easy to imagine ways to prevent this dynamical tunneling from happening, like creating uneven channels. Different injections could also be used to control the population of the branches~\cite{brandstotter2019shaping}.

\section{Discussion}
\label{sec:Discussion}

The results of this work connect and complete different areas in nonlinear dynamics. Varying the energy of classical trajectories and quantum mechanical wave packets in periodic potentials gives rise to multiple dynamical regimes. Different kinds of classical diffusion have been reported for values of the energy comparable or below the potential barriers~\cite{Cristadoro2014,Zarfaty2018,lorentzgas}.  At higher energies, typically several times larger than the height of the potential, we find the branched flow regime as demonstrated here for a periodic potential. In branched flow, individual trajectories fly over the potential and are barely affected by it, but successive interactions force the manifolds to fold onto themselves creating cusps and stable regions in phase space that give rise to the branches. Moreover, by using periodic potentials, the connection between classical chaos and branched flow has become evident.

The ideas presented here also lead to important questions in condensed matter physics. Branched flow is a transient regime in time and space, so electrons will eventually resemble Bloch waves. However, it may well be the case that this transient behavior dominates for very long distances and live for very long times. In particular, superwires demonstrated here can remain stable almost forever, except possibly for tunneling. Electrons traveling through these dynamical channels of the superlattices would have zero-resistivity. Although the persistence of the channels under perturbations still needs to be studied, it is hard to imagine how phonons could interact with electrons in these superwires.

\section{Acknowledgments}

We thank Alhun Aydin, Anatoly Obzhirov, Elizabeth Kozlov, Joonas Keski-Rahkonen, Shan Deneen, and Kobra Nasiri Avanaki for useful discussions. Financial support from NSF CIQM grant no. DMR-12313 19, and NSF CHE 1800101 is acknowledged. A.D. thanks the Real Colegio Complutense (RCC) that partially supported his research at Harvard University, and he also acknowledges support from the Spanish State Research Agency (AEI) and the European Regional Development Fund (ERDF, EU) under projects FIS2016-76883-P and PID2019-105554GB-I00.

\bibliography{biblio_branched}
\bibliographystyle{apsrev4-1}

\end{document}